\documentclass{jfm}
\usepackage{lineno,hyperref}
\usepackage{graphicx}
\usepackage{mathtools}
\usepackage{epstopdf, epsfig}
\usepackage{tikz}
\usepackage[ruled]{algorithm2e}
\usepackage{booktabs}
\usepackage{array}
\usepackage{amsmath}
\usepackage{multirow}
\usepackage{comment}
\usepackage{natbib}
\usepackage[]{color}
\usepackage[]{xcolor}
\usepackage{todonotes}
\usepackage{placeins}
\usepackage{relsize}
\usepackage{soul}
\usepackage{nomencl}
\usepackage{tcolorbox} 
\usepackage{float}
\usepackage{upgreek}

\newcolumntype{x}[1]{>{\centering\arraybackslash\hspace{0pt}}p{#1}}

\def\sl#1{\textcolor{black}{#1}}

\allowdisplaybreaks[1]
\shorttitle{Assimilation of wall-pressure measurements on a cone}
\shortauthor{D. A. Buchta, S. Laurence, \& T. A. Zaki}

\title{\Large Assimilation of wall-pressure measurements \\in high-speed flow over a cone}

\author{David A. Buchta\aff{1}
{Stuart J. Laurence}\aff{2}
 \and Tamer A. Zaki\aff{1} 
\corresp{\email{t.zaki@jhu.edu}}
}

\affiliation{\aff{1}\scriptsize Department of Mechanical Engineering, Johns Hopkins University, Baltimore, MD 21218, USA
\aff{2}Department of Aerospace Engineering, University of Maryland, College Park, MD 20742, USA}
 
\begin{document}

\maketitle

\abstract{
A nonlinear ensemble-variational (EnVar) data assimilation is performed in order to estimate the unknown flow field over a slender cone at Mach-6, from isolated wall-pressure measurements. The cost functional accounts for discrepancies in wall-pressure spectra and total intensity between the experiment and the prediction using direct numerical simulations (DNS), as well as our relative confidence in the measurements and the estimated state. We demonstrate the robustness of the predicted flow by direct propagation of posterior statistics. The approach provides a unique first look at the flow beyond the sensor data, and rigorously accounts for the role of nonlinearity unlike previous efforts that adopted ad-hoc inflow syntheses.  Away from the wall, two- and three-dimensional assimilated states both show rope-like structures, qualitatively similar to independent schlieren visualizations. Despite this resemblance, and even though the planar second modes are the most unstable upstream, three-dimensional (3D) waves must be included in the assimilation in order to accurately reproduce the wall-pressure measurements recorded in the Ludwieg-Tube facility. The results highlight the importance of three-dimensionality of the field and of the base-state distortion on the instability waves in this experiment, and motivate future measurements that probe the 3D nature of the flow field.}

%===========================
%   Introduction
%===========================
\vspace{-6pt}
\section{Introduction}
Hypersonic boundary-layer transition is extremely sensitive to environmental disturbances. 
Accurate transition predictions are therefore challenging in uncertain environments, especially when measurements are limited, for example for flight vehicles that are commonly instrumented with isolated wall-pressure probes. 
To reduce uncertainty, the current work is the first to infuse isolated wall-pressure measurements from a physical experiment in direct numerical simulations (DNS) of a Mach-6 boundary layer over a sharp cone. The computation thus reproduces the measurements and provides an unprecedented window into the transition process of the experiment.

Computational studies of transition on cones have focused on basic breakdown mechanisms,
and have sought to explore receptivity and to model the disturbance environment in experiments.
Simulations of fundamental and subharmonic resonant instability waves initiated by a wall forcing yield controlled breakdown scenarios that are suited for analysis~\citep{hader2019direct}, but they can deviate significantly from measurements of surface heat flux and wall pressure spectra \citep[e.g.~an order of magnitude difference in wall-pressure fluctuation magnitude,][]{Chynoweth2019history}.  A significant improvement was achieved by \cite{hader2018towards} who used a simple random-noise inflow forcing to model the receptivity processes from acoustic waves in a Mach-6 flow. 
However, significant over-predictions remain: the streamwise average heat flux by 65\% and peak wall pressure power spectra by over a factor of 20.
The above studies are valuable for understanding canonical transition scenarios and phenomenology from the experiments. The focus of the present work is to adopt a robust approach with objective guarantees that the simulations reproduce experimental measurements.

The herein adopted methodology systematically uses experimental measurements to rigorously determine the disturbance environment.  Our particular focus will be the experiments by \citet{kennedy2022characterization}, where wall-pressure measurements were recorded on a $7^\circ$ straight cone with a sharp nose in the Mach-6 Ludwieg tube facility at the Air Force Research Laboratory (AFRL).
The instrumentation layout is typical for this type of research, with pressure probes arranged in a streamwise ray to provide time-resolved information on streamwise amplification of instabilities; in addition, three azimuthal probes are placed to assess axi-symmetry of the pressure fluctuations.
The wall-pressure data indicate that the second mode feature prominently without full breakdown to turbulence.
Recent computations using the axisymmetric nonlinear parabolized stability equations (NPSE)
qualitatively capture the trend of the N-factor, without quantitative agreement~\citep{kennedy2022characterization} or guidance on how to objectively select the upstream disturbance spectra.
Furthermore, the role of three-dimensional waves and their relative amplitudes remains unknown since the simulations were axisymmetric.  The present approach will consider the relevant two- and three-dimensional instability waves and optimize their amplitudes using a variational framework so that their nonlinear evolution best reproduces the available measurements 
thereby establishing confidence in the entire reconstructed flowfield.
The following section details the experimental setup, numerical simulation, and data assimilation framework. Section \ref{sec:results} presents the outcomes of the data assimilation, including an analysis of the nonlinear dynamics from the reconstructed flow field that faithfully reproduces the measurements. Finally, a conclusion is provided in \S\ref{sec:conclusion}.

%===========================
%   Flow configuration
%===========================
\vspace*{-12pt}
\section{Flow configuration and methodology}
\label{sec:method}
%----------------
%   Experiment
%----------------
\subsection{Experiment and measurements}\label{sec:experiment}
\begin{figure}
    \centering
    \includegraphics[width=\textwidth]{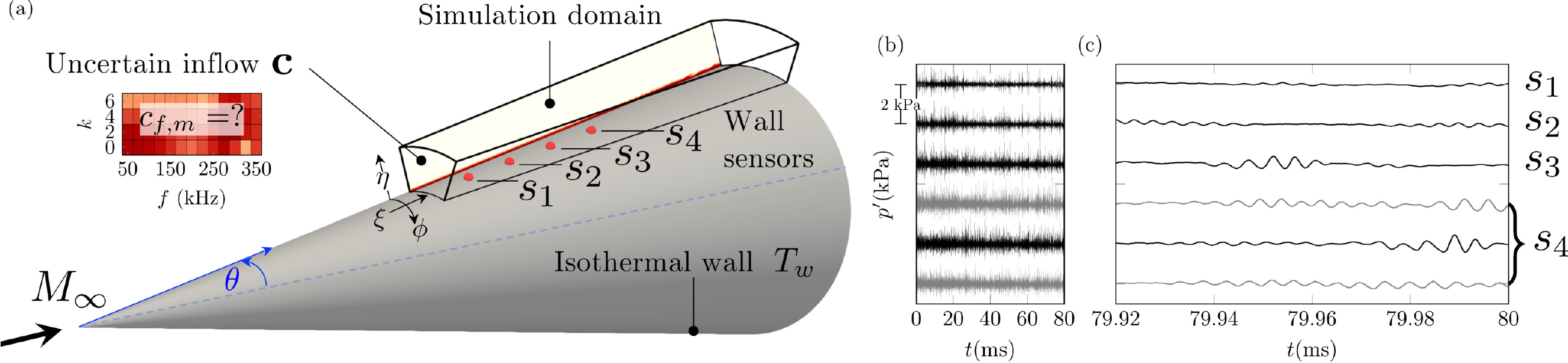}
    \caption{(a) Schematic of the flow configuration. (b) Extended time series of the PCB pressure data and (c) a detail for a 80~$\upmu\text{s}$ window. Signals are offset by 2 kPa for clarity.}
    \label{fig:configuration}
    \vspace{-6pt}
\end{figure}

The experimental flow configuration is shown in figure~\ref{fig:configuration}(a). 
The Mach $M_\infty=6.14$ flow developed as part of the dynamics of a Ludwieg tube, whereby heated ($T_0=450\,\text{K}$) and pressurized gas was contained in the charge tube upstream of the converging-diverging nozzle.  Once the fast-acting valve was opened, the gas accelerated through the nozzle to calculated free-stream conditions of $U_\infty=904\,\text{m/s}$, $T_\infty=54\,\text{K}$, and $\rho_\infty=0.0274\,\text{kg/m}^3$, yielding a unit free-stream Reynolds number $Re_\infty/L=7.11\times10^6\text{/m}$. The total test time was approximately 0.2 seconds, which was separated into two approximately time-stationary periods, lasting 100 milliseconds. Data from the second period is analyzed for this work \citep[for detailed characteristics of AFRL Ludwieg Tube, see][]{kimmel2017afrl, kennedy2022characterization}.

The test article was a $\theta=7^\circ$ half angle circular cone with a sharp nose (tip radius $r_n$=0.508\,\textrm{mm}) and a total length of 414\,mm. The model was installed at zero incidence to the streamwise direction; it was at room temperature ($T_w=300$ K) prior to the start of the experiments, and changes to the surface temperature during the brief test time are small enough that they can be ignored. The cone was instrumented with six PCB model 132A piezo-electric pressure sensors, positioned at four streamwise positions $\left\{s_1,s_2,s_3,s_4\right\}=\left\{215, 241, 266, 291, 316\right\}\,\text{mm}$, measured downstream of the nose. 
In addition, the $s_4$ row has two probes offset from the primary ray of sensors by $\pm 8.5^\circ$ in order to assess azimuthal dependence.  
Simulations and test data suggest that the effective sensing area has a diameter of approximately $d_s=0.97$ mm~\citep{ort2019influence}. For the case considered, time-resolved pressure data are available for six probes, acquired at a frequency of 5 MHz. A sample of the measurements is shown in figures~\ref{fig:configuration}(b,c). 
In (c), the pressure signatures show the presence of second-mode wave packets amplifying and advecting downstream. For any instance in (b,c), 
the signals do not appear to be chaotic with a broad range of time scales, suggesting the flow has not transitioned to turbulence at the sensors locations. 
High-speed schlieren measurements were also acquired, which provide additional points of validation above the wall. However, our focus is on wall data which are considered to be the primary modality taken during flight.

\begin{figure}
    \centering
    \includegraphics[width=\textwidth]{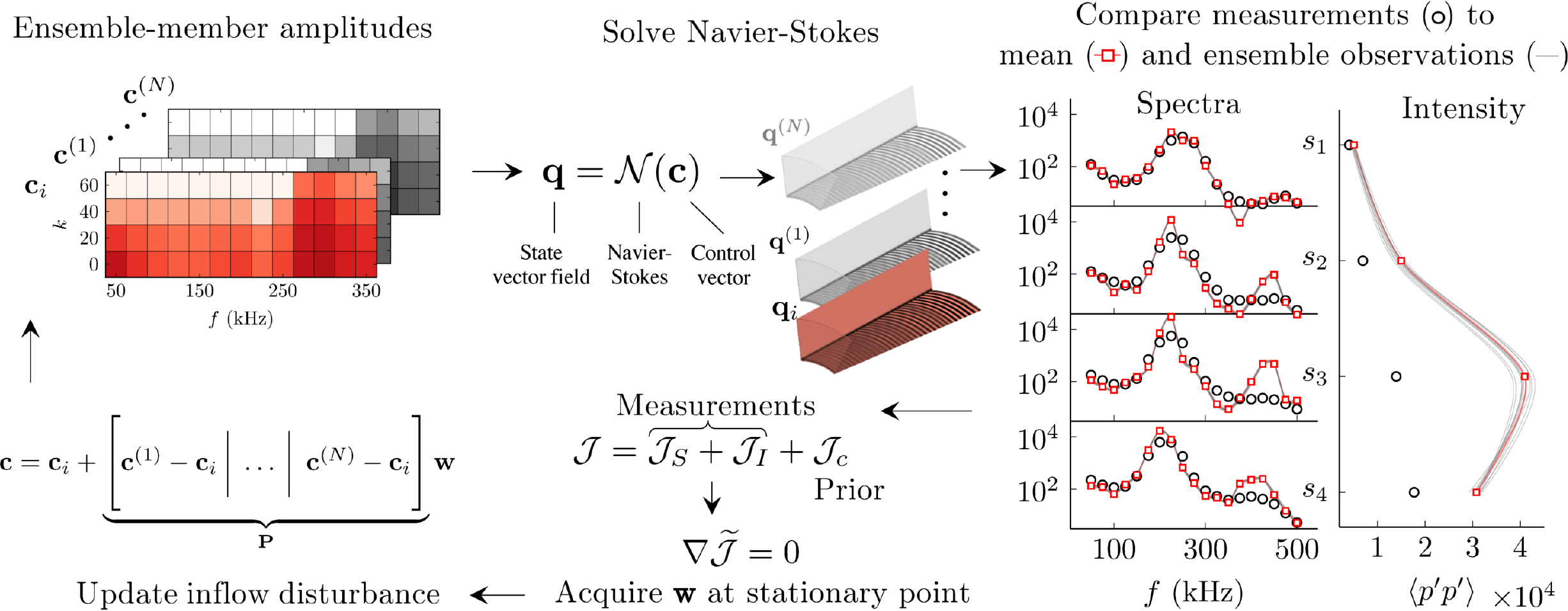}
    \caption{Schematic of the EnVar data assimilation framework. }
    \label{fig:envar:schematic}
    \vspace{-6pt}
\end{figure}

An analysis of the experimental data can aid in the choice of inflow frequencies and wavenumbers adopted in the data-assimilation simulations. The time-resolved pressure measurements are Fourier analyzed using the Welch method, similar to previous efforts~\citep{casper2016hypersonic,kennedy2022characterization}.  The window size of 80~$\upmu\text{s}$ (fig \ref{fig:configuration}c) furnishes $10^3$ non-overlapping Hann windows, which yields converged spectral amplitudes. 
The spectral resolution was halved by combining every two frequency bins, while conserving energy, in order to reduce the dimensionality of the data-assimilation problem. 
Although the flow is instantaneously three dimensional (figure \ref{fig:configuration}c), it is statistically axisymmetric under nominal conditions, i.e., zero incidence. 
Since the experimental setup was nominally axisymmetric, we assume homogeneity in the azimuthal direction and average the spectra of the three probes at $s_4$.
The post-processed measurements are then concatenated into two vectors,
\vspace*{-4pt}
\begin{align*}
    \mathbf{m}_S=\left[\hat{p}\hat{p}^\star(s_1,f_1), \ldots, \hat{p}\hat{p}^\star(s_4,f_{N_f})\right]^\top \text{and}\,\,\,\mathbf{m}_I=\left[{\sum}_i\hat{p}\hat{p}^\star(s_1,f_i), \ldots,{\sum}_i\hat{p}\hat{p}^\star(s_4,f_i)\right]^\top\!\!\!\!,
\vspace*{-12pt}
\end{align*}
representing the individual spectral amplitudes and overall intensity of the signal. The experimental spectra and signal intensity, used in the data assimilation, are shown with symbols in figure~\ref{fig:envar:schematic}. At position $s_1$, the spectrum is dominated by a peak between $f=225$-$275$ kHz, which is expected due to the presence of unstable second modes on straight cones~\citep{kennedy2022characterization}.
As the boundary layer thickens downstream, the frequency for the peak energy decreases. By $s_4$, the peak amplitude is near 200 kHz, and higher-harmonics, near 400 kHz, also appear, possibly due to nonlinearity from these high-amplitude waves. The total intensity of the signal amplifies between $s_1$-$s_3$ and appears to saturate between $s_3$ and $s_4$, also signaling the presence of nonlinear effects. The goal of this work then is to identify the unknown two- and three-dimensional instability waves upstream of $s_1$ that reproduce these measurements, and to study the associated flow field beyond the scope of the sensor data.

%----------------
%   Simulation
%----------------
\vspace*{-6pt}
\subsection{Simulation configuration}
The simulation domain is shown in figure~\ref{fig:configuration}. The flow behind the cone-generated shock is considered in order to interpret the measured wall-pressure spectra in terms of post-shock, boundary-layer disturbances. The base flow is axisymmetric and obeys the Taylor-Maccoll approximation
above the boundary layer, which is a solution to the Blasius equations in coordinates parallel ($\xi$) and normal ($\eta$) to the cone surface using the Mangler-Levy-Lees transformation.  
Prescribing the initial base state, $\mathbf{q}_B(\xi,\eta)$, in this manner is a well-established approach~\citep{sivasubramanian2015direct}. 
The inflow Reynolds number, based on the post-shock, boundary-layer-edge conditions ($U_e=880\,\text{m/s}$, $T_e=65\,\text{K}$, and $\rho_\infty=0.0458\,\text{kg/m}^3$) is $Re_o\equiv \rho_e U_e L_o/\mu_e =1345$ using the length scale $L_o = \sqrt{\nu_e \xi_o /U_e}$.  At this streamwise position, the inflow condition is expressed as a superposition of the base state and instability waves, 
\vspace*{-6pt}
\begin{align}\label{eq:inflowSuperposition}
    \mathbf{q}_o&=\mathbf{q}_B(\xi_0,\eta) + Re\left(\sum_m\sum_n c_{\scriptscriptstyle{{n,m}}} \widehat{\mathbf{q}}_{\scriptscriptstyle{n,m}}(\eta) \exp[\mathrm{i}k_m \phi-\mathrm{i}\omega_n t + \mathrm{i} \varphi_{\scriptscriptstyle{{n,m}}}] \right)\!,
\vspace*{-12pt}
\end{align}
where $c_{\scriptscriptstyle{{n,m}}}$ is the  amplitude, $\widehat{\mathbf{q}}_{\scriptscriptstyle{{n,m}}}$ is the discrete, slow-mode profile, and ($k_m$, $\omega_n$, $\varphi_{n,m}$) are the azimuthal wavenumber, frequency, and relative phase, respectively, for each $(n,m)$ instability wave.
The data assimilation attempts to identify the vector $\mathbf{c}=[\ldots,c_{\scriptscriptstyle{{n,m}}},\ldots]^\top$ 
of amplitudes of these waves at the inlet. We assume the average spectra is independent of relative modal phase ($\varphi_{n,m}$), which is reasonable given the random nature of free-stream tunnel forcing and the relatively long-time acquisition of the spectra. To reflect this assumption and lack of information especially for 3D waves, each phase is independently selected from a uniform random distribution $\varphi_{n,m}=\left[-\pi, \pi \right]$ and remains fixed during the course of the data assimilation. 
In contrast, if the objective were to assimilate the instantaneous wall-pressure signals, the relative phases of the inflow modes would need to be included in the control vector and accurately estimated~\citep{buchta_zaki_2021}.
\FloatBarrier
We consider 52 frequency and azimuthal wavenumber pairs that encompass Mack's first- and second-mode instabilities for the Reynolds-number range $1345 \leq \sqrt{Re_\xi} \leq 1872$. This size of the control vector was based on spectral analysis of the pressure-probes data to determine the dominant frequencies in the observations (see figure~\ref{fig:results}a) and preliminary testing.
The frequencies and integer azimuthal wavenumbers 
are $f\in \left[50,75,\ldots,350\right]\,\text{kHz}$ and $k\in \left[0,20,40,60\right]$, respectively. For each $(f,k)$-pair, we consider the most unstable slow modes~\citep{fedorov2011transition} over the Reynolds number range of interest, from the discrete Orr–Sommerfeld and Squire spectra. For the computation of the inflow instability profiles $\widehat{\mathbf{q}}_{\scriptscriptstyle{{n,m}}}(\eta)$ only, the effect of curvature was neglected, i.e.\,we adopt the wavenumber $\beta_m=k_m / r_o$ where $r_o$ is the radius of the cone at the inflow, which is a reasonable assumption for this Reynolds number~\citep{malik1991stability}. 
Once the inflow condition (\ref{eq:inflowSuperposition}) is prescribed, the downstream evolution in the DNS/LNS accounts for curvature fully.
Since the experimental measurements lack three-dimensional information to guide the selection of relevant oblique modes, we consider a range of azimuthal wavenumbers that are linearly unstable ($k\leq 60$), and discretize this range in a manner that we can efficiently search and that enables the generation of important higher harmonics by nonlinear interactions.  Beyond the inflow, the azimuthal discretization of the computations permits the formation of waves up to $k=540$, thus providing ample resolution for secondary instabilities that arise due to fundamental resonance between oblique and planar instability waves.

Most of the present simulations solve the compressible, nonlinear Navier-Stokes equations in curvilinear coordinates (termed DNS) for the simulation domain shown in figure~\ref{fig:configuration}. Linearized dynamics are acquired with the exact same solver assuming a steady base state and neglecting nonlinear interactions only (termed LNS). LNS is used to provide an initial estimate of the unknown inflow spectra that will be refined to account for nonlinear effects; LNS will also be referenced in the discussion to contrast the disturbance field to the nonlinear evolution.   The gas is assumed ideal with ratio of specific heats $\gamma= 1.4$ and temperature-dependent viscosity, following a power-law formula, $T^n$, where $n$ is determined from measured viscosity data between the boundary-layer edge and wall temperatures ($T_e=65$\,\text{K} and $T_w=300$\,\text{K}). The flow equations are solved using a standard fourth-order Runge–Kutta scheme and interior fourth-order finite differences. Near boundaries, stencils are biased and accuracy is reduced to second order. Second derivatives in the viscous terms are evaluated using repeated first derivatives, instead of discretizing the operator directly~\citep{mattsson2004summation}, which requires adding high-order numerical dissipation of short waves to stabilize the solution \citep[for details, see][]{vishnampet2015practical}.
The computational domain size $(L_\xi, L_\eta,L_\phi)$=(105\,\text{mm}, 17.6\,\text{mm}, $36^\circ$) is discretized with $(N_s, N_y, N_\theta)=(751, 201, 108)$ grid points.  The azimuthal size was chosen to extend beyond the peripheral probes at $s_4$ in order reduce any artificial correlation by the periodic boundaries. 
In viscous units, the grid spacing along the wall at the inflow is ($\Delta \xi^+$, $\Delta \eta^+$, $r_o \Delta \phi^+$) = (2.8, 0.1, 3.1). The azimuthal and streamwise grids are uniform, and the wall-normal grid spacing is stretched according to the transformation used in \citet{pruett1995spatial}.

For all of the simulations, the flow develops for 1.75 flow-through times, based on the edge velocity $U_e$, before data acquisition commences.  During 40~$\upmu\text{s}$, wall pressure is recorded at every time step; the time step size is chosen to ensure the $\text{CFL}\approx 0.4$ throughout the simulated time horizon. The finite-sensing area of the PCB probes is modeled by averaging the wall pressure, $\widetilde{p}_j = 1/N\sum_{i=1}^N p_i$, for $N$ grid points that satisfy the inequality $(x_j-x_i)^2+(y_j-y_i)^2+(z_j-z_i)^2 \leq (d_s/2)^2$ where $d_s$ is the sensing diameter. To take advantage of statistical homogeneity in the simulation, this operation is performed at the streamwise sensor position for all azimuthal grid points. The simple averaging used to model the sensing area in DNS was shown to improve adherence to measurements of pressure spectra and intensity~\citep{huang2020direct}. The data from the simulated probe pressure $\widetilde{p}$ are processed and concatenated into two vectors in the same manner as the experimental measurements,
\vspace*{-4pt}
\begin{align*}
    \mathbf{h}_S=\left[\hat{p}\hat{p}^\star(s_1,f_1), \ldots, \hat{p}\hat{p}^\star(s_4,f_{N_f})\right]^\top \text{and}\,\,\,\mathbf{h}_I=\left[{\sum}_i\hat{p}\hat{p}^\star(s_1,f_i), \ldots,{\sum}_i\hat{p}\hat{p}^\star(s_4,f_i)\right]^\top\!\!\!\!.
\vspace*{-12pt}
\end{align*} 
These vectors are used as inputs for the following data assimilation framework.

%-----------------------
%   Data assimilation
%-----------------------
\vspace*{-4pt}
\subsection{Data assimilation procedure}
The unknown amplitudes $\mathbf{c}$ of instability waves are sought using an ensemble variational framework~\citep{buchta_zaki_2021}. To initiate the nonlinear optimization, an estimate for the control vector $\mathbf{c}_0$ is determined using linear dynamics of the flow. This choice provides a more physics-based approximation than adopting white noise across the spectrum and, as such, a more accurate starting estimate for the subsequent nonlinear optimization. For the linear estimate, the control vector is computed by direct differentiation of
\vspace*{-4pt}
\begin{equation}\label{eq:lin:cost}
        \mathcal{J}_l=\frac{1}{2}||\mathbf{m}-\mathbf{L}\mathbf{v}||^2 + \frac{\varrho}{2}||\mathbf{v}||^2 
% \vspace*{-4pt}
\end{equation} 
with respect to the weights and identifying $\mathbf{v}$ at a stationary point $\nabla \mathcal{J}_l = 0$. Each column of matrix $\mathbf{L}=\left[\ldots~~|~~\mathbf{l}_{f,k}~~|~~\ldots\right]$ represents the wall observations acquired from an independent evolution of $i$-th instability wave at ($f,k$) using the linearized flow equations and an axisymmetric Blasius base state. Since the performance of the linear method is predicated on the validity of a linear assumption, only measurements from the first sensor position $s_1$, where harmonics of the large-amplitude waves are absent, are included for the estimate of $\mathbf{c}_0$. The second term in (\ref{eq:lin:cost}) represents a penalty which has two desired effects: (i) it mitigates against high-energy inflow  disturbances, and (ii) it improves the conditioning of the inverse problem. The value of $\varrho$ is chosen such that $\mathcal{J}_l/\mathcal{J}_{\mathbf{v}=0}=10^{-4}$, which ensures that the measurements are accurately reproduced to within $1$\% (see \S\ref{sec:results}). If the experimental conditions only trigger a linear boundary-layer response, then minimizing (\ref{eq:lin:cost}) will be sufficient to accurately assimilate the measurements and predict the inflow amplitudes. 

For the experiment considered in \S\ref{sec:experiment}, signatures of nonlinearity are present in the measurements. As a result, the initial estimate $\mathbf{c}_0$ of the control vector based on (\ref{eq:lin:cost}) is not sufficient, and the data assimilation requires an iterative, nonlinear optimization.  We consider the following cost,
\vspace*{-4pt}
\begin{equation}\label{eq:nonlinear:cost}
       {\mathcal{J}}={\frac{1}{2}||\log_{10}\mathbf{m}_S-\log_{10}\mathbf{h}_S||_{\Sigma_m^{-1}}^2}+{\frac{1}{2}||\mathbf{m}_I-\mathbf{h}_I||_{\Sigma_n^{-1}}^2}+\frac{1}{2}||\mathbf{c}-\mathbf{c}_i||_{\Sigma_c^{-1}}^2,
\vspace*{-4pt}
\end{equation}
which comprises three parts.
The first term is the logarithm of the pressure spectra. Since the spectra span several orders of magnitude, the logarithm ensures that the cost function is not solely focused on the spectral peak but rather targets the entire range of frequencies. The second term captures the importance of the spectral peak in determining the overall pressure intensity.
The final term represents the degree of trust in a prior control vector ($\mathbf{c}_i$) to avoid large steps from the previous control vector. The choice of the cost function can affect performance of the data assimilation, and our experience has shown that (\ref{eq:nonlinear:cost}) furnishes better accuracy in fewer iterations relative to, for example, pressure spectra on a linear scale. 

An ensemble variational (EnVar) technique is used to update the control vector in order to minimize the cost function~(\ref{eq:nonlinear:cost}). Figure~\ref{fig:envar:schematic} shows a schematic of the algorithm. An updated estimate of the control vector $\mathbf{c}$ is sought from its previous value $\mathbf{c}_i$ using a weighted superposition of ensemble members $\mathbf{c}^{(j)}$, specifically $\mathbf{c}_{i+1}=\mathbf{c}_i + \mathbf{P} \mathbf{w}$, where $\mathbf{P}=\left[\ldots~~|~~ \mathbf{c}^{(j)}-\mathbf{c}_i~~|~~\ldots \right]$ is a matrix of the perturbations vectors.
Each control vector, $\mathbf{c}_i$ and $\mathbf{c}^{(i)}$, is evolved with the governing equations, in this case the Navier--Stokes equations, to produce the spatio-temporal representation of the state $\mathbf{q}_i$ and $\mathbf{q}^{(j)}$.  
Model observations are then extracted from the mean and ensemble states using $\mathbf{h}_i=\mathcal{M}(\mathbf{c}_i)$ and $\mathbf{h}^{(j)}=\mathcal{M}(\mathbf{c}^{(j)})$; they are then compared with the experimental data in the cost function.  An estimate of the local gradient and Hessian furnish the optimal weights $\mathbf{w}$ by assuming a stationary point ($\nabla\widetilde{\mathcal{J}}=0$). 
Multiple updates of the control vector can be performed in the descent direction, until the cost stagnates or starts to increase. At this point, the entire EnVar process is repeated, iteratively and till convergence. The number of ensemble members in the main algorithm was $N_e=10$.  Each EnVar iteration thus involves $N_e+1=11$~simulations to acquire the local gradient, where the additional DNS corresponds to the mean of the ensemble. 
Previous characterizations of the performance of this method are provided elsewhere~\citep{mons2021ensemble,jahanbakhshi_zaki_2019,buchta_zaki_2021}. 
The results of the linear and nonlinear optimization are presented in the next section. 

%=======================
%   RESULTS
%=======================
\vspace*{-12pt}
\section{Results}\label{sec:results}
\begin{figure}
    \centering
    \includegraphics[width=0.9\textwidth]{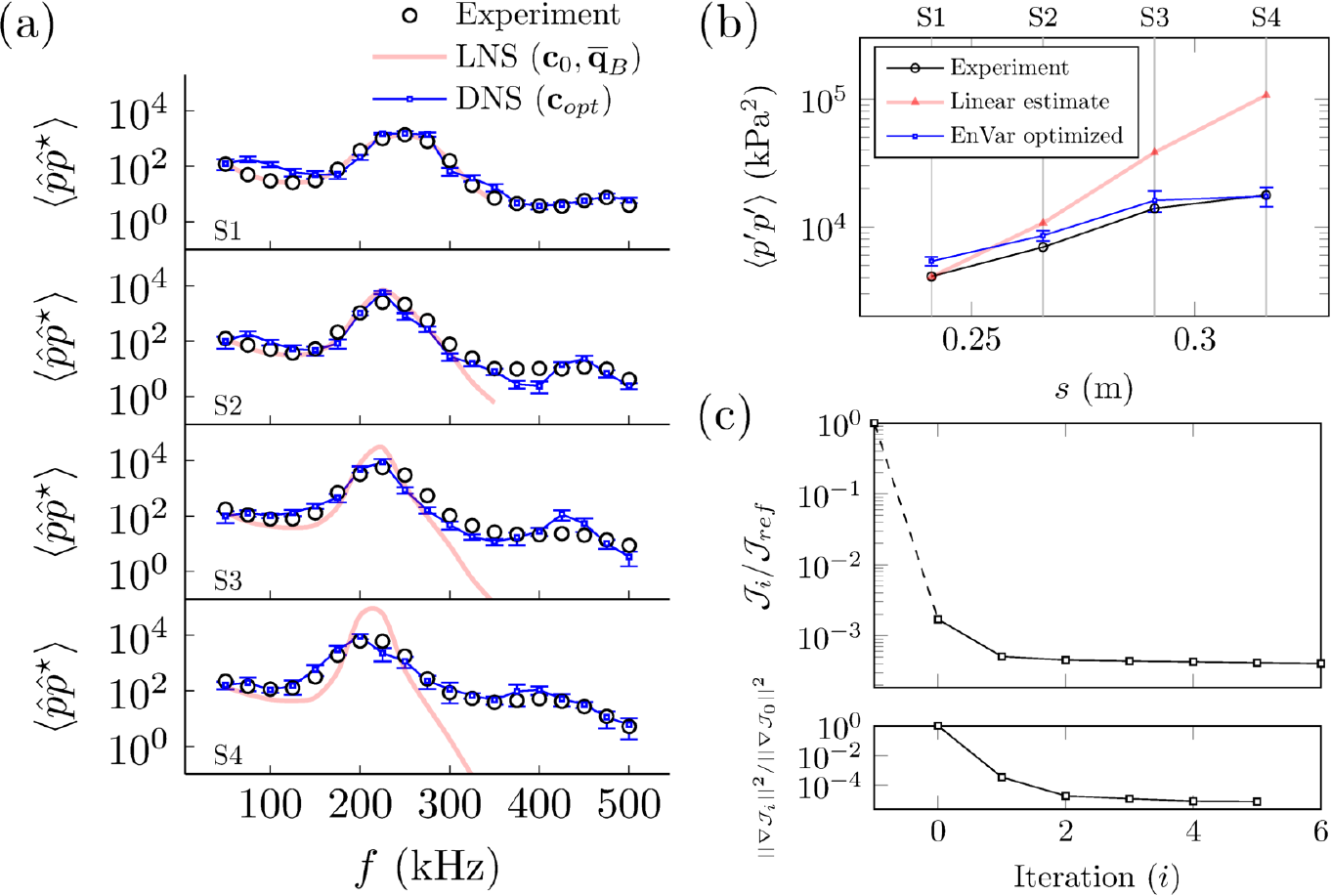}
    \vspace*{-8pt}
    \caption{(a) Power spectra at measurement stations for the experiment (circles), EnVar prediction (connected squares), and linear prediction (line).  (b) Integral of power spectra. The confidence in each predicted observations is represented by $\pm 75\sigma_i$ where $\sigma_i$ is the ensemble standard deviation of each i-th observation.  (c) Cost and gradient during data assimilation.}
    \label{fig:results}
    \vspace{-10pt}
\end{figure}
The outcomes of the linear and nonlinear assimilations are shown in figure~\ref{fig:results}. By design of (\ref{eq:lin:cost}), the linear estimate accurately approximates the experimental measurements at $s_1$ in (a) and (b).
However, the subsequent downstream predictions deviate from the measurements. The linear amplification of the second mode atop the Blasius base state is more intense than in the experiments, and modes with $f \gtrsim 300\,\text{kHz}$ decay rather than amplify. 
The linear dynamics do not reproduce the wider experimental spectra downstream or the accumulation of energy in frequencies adjacent to the energetic second modes. 
Absence of nonlinear interactions contributes to these discrepancies, specifically due to the omitted generation of higher harmonics and the lack of base-flow distortion, which will both be discussed below. 
Most importantly, these effects were also omitted from the linear estimate of the inflow disturbance spectra, and the disagreement with downstream measurements underscores the importance of a nonlinear interpretation of the observations.  

Nonetheless, figure~\ref{fig:results} demonstrates that our physics-based linear estimate of the inflow is a better initial guess for the nonlinear assimilation than an \emph{ad hoc} approach to inflow synthesis.
Starting from the linear estimate, figure~\ref{fig:results}(c) shows the cost reduction by nonlinear EnVar iterations relative to the LNS, $\mathcal{J}_{ref}$, and the decrease in the gradient magnitude.  After six iterations, the cost and gradient decrease over three and four orders of magnitude, respectively, and the trends indicate an optimum has been identified. Additionally, the EnVar approach enables propagated uncertainty of the prediction. By evolving ensemble members at the final iteration, the sample standard deviation for each observation, $\sigma_i$, is computed. These error bars, $\pm 75\sigma_i$ in figures~\ref{fig:results}(a,b), confirm the low uncertainty for our estimates of the measurement. A positive bias, not explained by parametric uncertainty, remains in the lowest amplitude signals at $s_1$ and $s_2$. Overall, however, the adherence of the DNS to the observations in terms of spectral amplitude and the overall intensity is encouraging. This level of agreement surpasses in accuracy what we were able to achieve using the same optimization procedure and considering 2D modes only. In fact, when the inflow is restricted to axisymmetric waves only (not shown), the objective function was 60\% higher, and the mismatch with the measurements was most egregious on the final sensor position.

\begin{figure}
    \centering
    \includegraphics[width=1\textwidth]{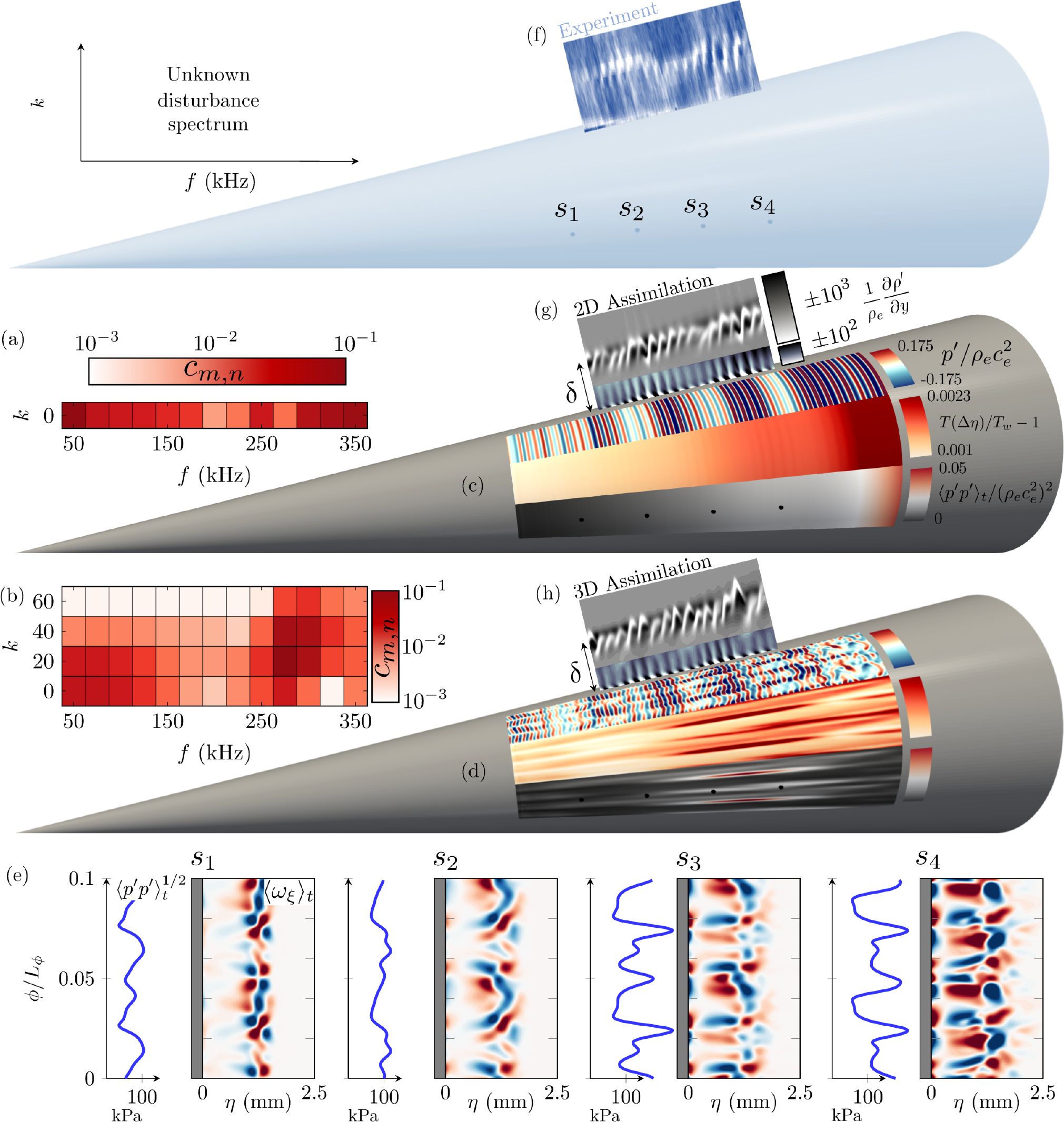}
    \vspace{-10pt}
    \caption{(a,b) Amplitudes of the optimized inflow spectra for the 2D and 3D assimilation. (c,d) Reconstructed wall quantities: instantaneous $p'$, time-averaged near-wall temperature and pressure intensity. (e) Spanwise variation of time-averaged wall pressure r.m.s and streamwise vorticity at sensors $s_1$-$s_4$. (f) Experimental schlieren and (g,h) numerical schlieren $\partial_y \rho'$. }
    \label{fig:viz}
\end{figure}

Figures~\ref{fig:viz}~(a) and (b) show the distribution of modal amplitudes predicted by the nonlinear data assimilation procedure, assuming 2D and 3D incoming disturbances, respectively. The axisymmetric reconstruction, which was less effective in reducing the cost function, requires three times more inflow energy in the two-dimensional waves in order to approximate the measurements with lower fidelity.  
The better-performing 3D interpretation prominently features the oblique waves ($275\leq f \leq300$ and $20\leq k\leq 40$), which demonstrates that three-dimensionality is required to reproduce experimental measurements even when the flow is non-turbulent within the sensing region\textemdash an important point that is often overlooked in the interpretation of wall-pressure data in high-Mach number transitional flows in terms of the dominant planar instability waves.

\begin{figure}
    \centering
    \includegraphics[width=1\textwidth]{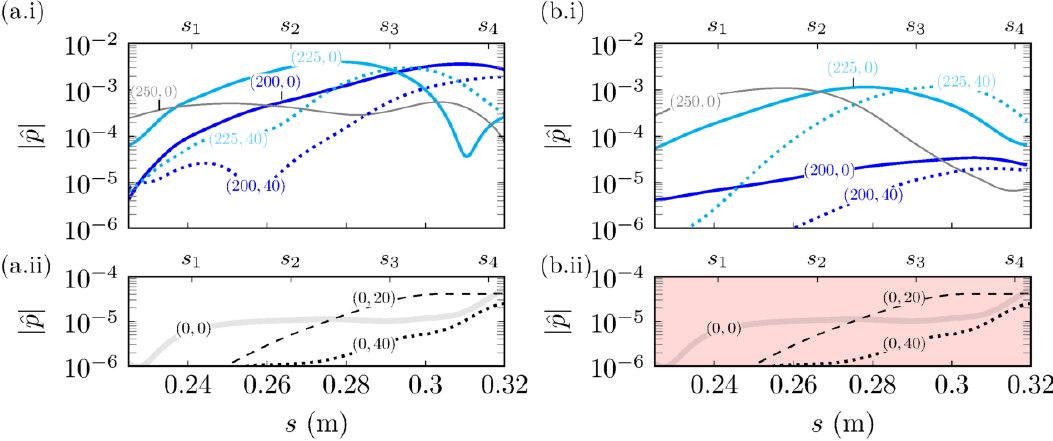}
    \vspace{-12pt}
    \caption{Streamwise evolution of wall-pressure spectra. Dominant unsteady modes in (a.i) DNS and (b.i) LNS with three-dimensional distorted base state $\mathbf{q}_B=\langle \mathbf{q}\rangle_t$.  
    (a.ii) Dominant steady modes from DNS, which are part of the distorted base state used in LNS (reproduced in (b.ii)). }
    \label{fig:wallSpectra}
    \vspace{-12pt}
\end{figure}

In addition to discovering the unknown inflow that best reproduces the measurements, figure~\ref{fig:viz} shows that the data assimilations identify the entire time-dependent and three-dimensional flow field, far beyond the measurements probes. 
The instantaneous pressure fluctuations in (c,d) along the wall reveal  long-streamwise-wavelength, two-dimensional envelopes of instability waves
distributed along the cone.  Despite its appearance, we verified using spectral analysis (c.f. figure~\ref{fig:wallSpectra}) that this pattern is not an amplitude modulation.  Instead, the pattern of repeated amplification and decay is due to an interference of waves with approximately the same advection speed. Atop this pattern, the signature of oblique fluctuations is prominent in (d), where the instantaneous pressure fluctuations are more realistic than in the 2D assimilation since they more closely reproduced the measured wall-pressure intensity.  Figure~\ref{fig:viz}(d) also shows the time-averaged near-wall temperature and squared-pressure fluctuations, both of which feature streamwise-elongated, intense streaks extending across multiple sensor positions.

\begin{figure}
    \centering
    \includegraphics[width=\textwidth]{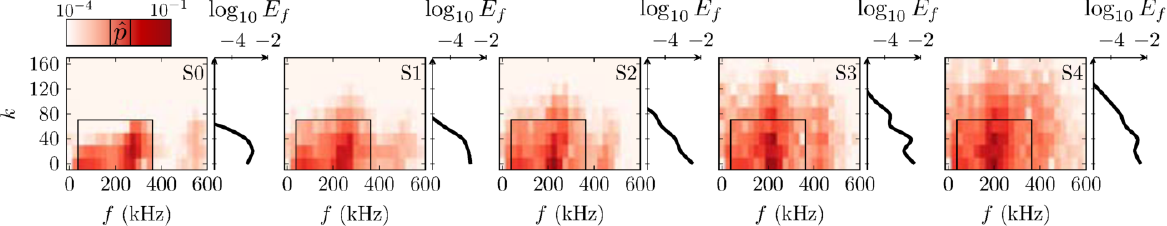}
    \vspace{-12pt}
    \caption{Two-dimensional spectra of wall pressure at the inflow ($s_0$) and probe locations ($s_1-s_4$). Curves are integrals of energy over frequency, $E_f(k)=\int \hat{p} \hat{p}^\star\,df$.}
    \label{fig:wallspectra}
\vspace{-12pt}
\end{figure}

Besides the fact that the 3D interpretation is most accurate, the three-dimensional characteristics of the flow share common features observed in previous DNS of fundamental resonance breakdown ~\citep{sivasubramanian2015direct,hader2019direct} and random-noise forcing~\citep{hader2018towards}, e.g.~steady streamwise elongated near-wall hot streaks and pairs of counter-rotating vortical structures. Thus, the reconstructed 3D inflow produces previously realizable flow fields, but most importantly quantitatively agrees with measurements. 
Figure~\ref{fig:viz}(e) focuses on the three-dimensionality of the wall-pressure intensity and the streamwise vorticity above the sensor positions. At $s_3$ and $s_4$, prominent wall-pressure intensity streaks, with peak amplitude nearly twice the mean, fall between pairs of intense counter-rotating vortices above which fluid is ejected upward. 
These structures correspond to a distortion $(f,k)=(0,20)$ to the base flow, which can support parametric resonance (subharmonic, fundamental or detune) with the instability waves\textemdash a point that we will revisit below.

We recall that independent schlieren measurements were performed in the experiment, and were kept for blind comparison to the outcome of assimilating the wall-pressure data.  Figure~\ref{fig:viz}(f) shows an instantaneous realization from the experimentally acquired schlieren data, and captures the rope-like structures near the boundary-layer edge.  
Figures~\ref{fig:viz}(g) and (h) show numerical schlieren from both the 2D and 3D assimilated flows.  Both (g,h) feature the rope-like structures in the wall-normal density gradient, bear clear similarity to the experimental schlieren images in that region, and are similar to one another near the wall.  This qualitative agreement serves as another note of caution against the often-adopted two-dimensional interpretation of pre-transitional experimental schlieren measurements.  In the present case, only the three-dimensional inflow disturbance field can justify, and reproduce, the measured wall-pressure data.

We initially argued that the experimental measurements show symptoms of nonlinearity since harmonics of the dominant frequencies were observed in the spectra (figure \ref{fig:results}), and subsequently that three-dimensionality is important for successful interpretation of the data. In order to assess the extent of nonlinearity and three-dimensionality in the 3D assimilated state, in figure \ref{fig:wallspectra} we report the wall-pressure spectra as a function of $(f,k)$, evaluated at the inflow ($s_0$) and at the four streamwise sensor positions ($s_1$-$s_4$).  The broadening of the spectra starting at $s_{3}$ coincides with the prominent base-state distortion in figure \ref{fig:viz}(d), which is an important nonlinear effect.  The figure also shows the energy integrated over the entire frequency range and plotted versus the azimuthal wavenumber. The energy in the oblique waves is commensurate with that in planar ones at $s_{\{1,3,4\}}$, i.e.~throughout the majority of the domain.

Select modes of the wall-pressure spectra from DNS are reported in figures \ref{fig:wallSpectra}(a.i, a.ii), separated into unsteady waves $(f,k)$ and base-state distortions $(0,k)$.  For comparison, figure \ref{fig:wallSpectra}(b.i) shows the linear evolution of the unsteady modes; For the base flow in the LNS, we adopted the time-averaged, three-dimensional, distorted state computed in the DNS, and as such we reproduce figure \ref{fig:wallSpectra}(a.ii) in (b.ii).
The spectra highlight the necessity of nonlinear assimilation of the measurements. 
Between the inflow and $s_1$, nonlinearity causes the spectra in (a.i) to already differ from the linearized case (b.i).  For example, linear theory predicts that $(f,k)=(250,0)$ is important at $s_1$, but this interpretation would lead to an under-prediction at the inflow; the nonlinear evolution of the mode, which is relatively muted than amplified, would not match the measurements.  Another example relates to the dominant planar waves (225,0) and (200,0) from the DNS, which are respectively delayed and under-predicted in LNS.  To compensate, a linear interpretation of the measurements would require large, practically nonphysical, inflow amplitudes for these second-mode instabilities; such prediction would lead to drastically poor nonlinear flow response.

As for the the three-dimensional waves, the most energetic mode near $s_3$ is $(f,k)=(225,40)$. This instability has similar amplification in subfigures (a.i) and (b.i), and this agreement was only possible by adopting the nonlinearly distorted base flow in the LNS.
The growth of this mode also parallels the trend of the base-state distortion $(0,20)$.  The latter corresponds to the streaks in wall-pressure intensity in figure \ref{fig:viz}(d), and can support parametric resonance leading to the amplification of $(f,20 m)$, where $m=1,2,...$ for fundamental resonance.
The most dominant of these waves, namely $(f,k)=(225,40)$, would elude detection by the experimental azimuthal probes at $s_4$, whose Nyquist wavenumber is $\approx 21$.

In summary, for the considered experiment, a linear interpretation even of the most upstream data does not reproduce the downstream measurements, and a focus on two-dimensional instability waves is fraught with uncertainty, even within the non-turbulent region. 
Only nonlinear assimilation of the wall-pressure sensor data provides rigorous prediction of the inflow amplitudes, and the present work demonstrated the importance of three-dimensionality.

\vspace{-14pt}\section{Conclusion}
\label{sec:conclusion}

The unknown upstream disturbances for a Mach-6 flow over a sharp cone were reconstructed by assimilation of spectral and statistical data from discrete wall-pressure probes. 
The data assimilation considered a multi-objective cost functional: (i) the logarithm of the pressure spectra promoted matching the data across a wide range of measured frequencies
and (ii) the integral of the spectra promoted the prediction accuracy of modes comprising most of the intensity. 
Fidelity of reproducing the measurements from the present Ludwieg-tube experiment is possible only when three-dimensional nonlinear interactions are incorporated in the assimilation framework. 
The need to reproduce these interactions accurately when considering more consequential and sensitive dynamics, e.g., transition to turbulence, remains of high importance.
For the present configuration, future measurements must probe three-dimensionality because the nonlinear optimization showed that only with 3D effects that we can reproduce the data. 

The data assimilation approach is robust, and can be applied to other experimental conditions.  If the dynamics are linear, the initial linear estimate of the control vector is sufficient to reproduce the measurements.  When important effects are not simulated, the quantitative deviation from the measurements can guide the improvement of the computational model, in our case by taking into account nonlinearity and three-dimensional instability waves.  While in this work we focused on predicting the upstream instability waves in the boundary layer, our methodology can also be applied to estimate incident free-stream disturbances, either post- or pre- leading-edge shock, including in the considerably uncertain disturbance environment faced in flight.

\par\bigskip
\noindent
\textbf{Acknowledgements}
The authors are grateful to the Air Force Research Laboratory (AFRL) for providing the measurements that were used in the present study.

\par\bigskip
\noindent
\textbf{Distribution Statement A}: Approved for Public Release; Distribution is Unlimited. PA\# AFRL-2022-2226.

\par\bigskip
\noindent
\textbf{Funding.}  This work was supported in part by the US Air Force Office of Scientific Research (grant no.\,FA9550-19-1-0230) and by the Office of Naval Research (grant no.\,N00014-21-1-2148). Computational resources were provided by the Maryland Advanced Research Computing Center (MARCC). 

\par\bigskip
\noindent
\textbf{Declaration of interests.} 
The authors report no conflict of interest.
 
\par\bigskip
\noindent
\textbf{Author ORCIDs.} \\
Tamer A. Zaki, \url{https://orcid.org/0000-0002-1979-7748}

\bibliography{references}
\bibliographystyle{jfm}

\end{document}